\documentclass[twocolumn, showpacs, preprintnumbers, amsmath, amssymb, superscriptaddress, pra]{revtex4}
\usepackage{graphicx}
\usepackage{dcolumn}
\usepackage{bm}
\usepackage{epsf}
\usepackage{amssymb}
\DeclareGraphicsExtensions{.eps}
\begin{document}

\title{ Correlated-Photon Imaging with Cancellation of Object-Induced Aberration}
\vskip4pc

\author{D.S. Simon}
\affiliation{Dept. of Electrical and Computer Engineering, Boston
University, 8 Saint Mary's St., Boston, MA 02215}
\affiliation{Photonics Center, Boston
University, 8 Saint Mary's St., Boston, MA 02215}
\author{A.V. Sergienko}
\affiliation{Dept. of Electrical and Computer Engineering, Boston
University, 8 Saint Mary's St., Boston, MA 02215}
\affiliation{Photonics Center, Boston
University, 8 Saint Mary's St., Boston, MA 02215}
\affiliation{Dept. of Physics, Boston University, 590 Commonwealth
Ave., Boston, MA 02215}

%
%


\begin{abstract} We show that a recently discussed apparatus
for aberration-cancelled interferometry may be modified to perform
correlated-photon imaging in the so-called "ghost" imaging configuration.
For objects in the vicinity of a particular
plane, the images are free of object-induced phase distortions.
This apparatus has the distinctive feature that it may be used to
superimpose images of two objects in a manner that could lead to useful effects and applications.
We show that the apparatus works using either quantum-entangled or classically correlated light sources.
\end{abstract}


\maketitle

\section{INTRODUCTION} Correlated-photon imaging, sometimes known as "ghost" imaging, was first discovered
using entangled photon pairs \cite{pittman} from spontaneous parametric downconversion (SPDC).
It has since been found
that most aspects of ghost imaging can be simulated using
spatially-correlated classical light \cite{bennink1, bennink2},
including thermal and speckle sources \cite{gatti, cai, valencia,
scarcelli, ferri, zhang}.
A separate line of research has shown that entangled photon pairs from
downconversion may also be used
to cancel some of the effects of frequency dispersion
\cite{steinberg, franson, minaeva1} or spatial dispersion
(aberration) \cite{bonato1, bonato2, simon1}. In \cite{simon1}, it
was pointed out that it is possible to construct an interferometer
such that if an object is placed in a particular plane then the
effects of all phase shifts induced by that object, including all
object-induced aberrations, will cancel in the resulting
coincidence rate. The goal here is
to move away from interferometry and to produce an analogous
effect in an imaging system. We show that this may be achieved by
a simple variation of the traditional ghost imaging apparatus of
figure 1. It is thus possible to produce
images of the object's amplitude transmittance
profile, undistorted by phase effects as long as the
object is entirely contained within a small region near the
special plane mentioned above.
(For simplicity we will only discuss transmission here;
the case of reflection at the object is similar.)
We then show that, although an entangled source was
required for the temporal correlation experiments discussed in
\cite{bonato1, bonato2, simon1}, a classical source with
transverse spatial correlation will suffice for imaging.

In addition, if {\it two} object are placed in the resulting optical system, one in each arm, the image produced will simply be the point-by-point product of the images that would be generated by each of the two separately. This is a new feature that does not appear if two objects are placed in the arms of other types of ghost imaging systems. We will comment on several possible applications of this effect below.

\begin{figure}
\centering
\includegraphics[totalheight=2.0in]{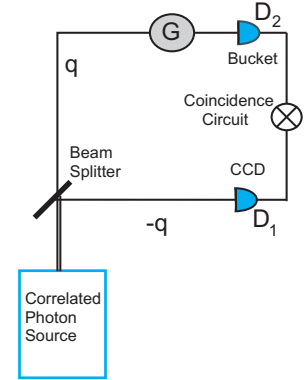}
\caption{\textit{(Color online) Schematic depiction of correlated-photon imaging
setup. The photons in the two arms have anticorrelated transverse
momenta $\pm {\mathbf q}$.}}
\end{figure}

We begin by briefly reviewing ghost imaging in section 2, followed by a review of aberration-cancelled
interferometry in section 3. We then show how a small change converts the
aberration-cancelled interferometer into a new type of ghost imaging
system. We analyze this imaging system first with an entangled
light source in section 4, then with a classical source in section 5. Finally, in section 6 we discuss an
important technical point about the need for lenses in front of
the detectors, followed by conclusions in section 7.


\section{CORRELATED-PHOTON IMAGING} Correlated-photon imaging
or "ghost" imaging \cite{pittman} is done with
an apparatus like the one depicted schematically in figure 1. In
the original version, the correlated photon source is a
$\chi^{(2)}$ nonlinear crystal pumped by a laser, leading to SPDC.
Entangled photon pairs with anticorrelated momentum components
${\mathbf q}$ and $-{\mathbf q}$ transverse to the propagation
direction travel along the two arms of the apparatus. The object
to be viewed is placed in arm 2 (the upper branch), followed by a
bucket detector, $D_2$. $D_2$ can not record any information on
the position or momentum of the photon that reached the object; it
can only tell us whether or not the photon reached the detector
unimpeded by an object. The other arm has no object, and all the
photons reach a CCD camera or array of pointlike detectors without
hindrance. A lens may be inserted in this branch for image formation.
A coincidence circuit is used to record a count every
time a photon detection occurs simultaneously (within a short time
window) at each detector. By plotting the coincidence rate as a
function of position ${\mathbf x_1}$ in detector 1, we build up an
image of the object. This is true even though photons that
actually encountered the object in branch 2 left no record of the
object's position, and the photons in branch 1 that do carry
position information never encounter the object.

The crucial ingredient is the spatial correlation of the photon
pair. It was found \cite{bennink1, bennink2} that entanglement was
unnecessary: a classical source with anticorrelated transverse
momenta could mimic the effect. The correlated light source in
this case consists of a beam steering modulator (a rotating
mirror, for example) directing a classical light beam through a
range of ${\mathbf q}$ vectors, illuminating different spots on
the object. The beamsplitter turns the single beam of transverse
momentum ${\mathbf q}$ into a pair of beams with momenta ${\mathbf
q}$ and $-{\mathbf q}$.  The results were similar to those with
the entangled source, but with half the visibility. It was later
shown that thermal and speckle sources may also lead to ghost
imaging (\cite{gatti, cai, valencia, scarcelli, ferri, zhang}).

\section{SUMMARY OF ABERRATION CANCELLATION IN QUANTUM INTERFEROMETRY} Consider the setup shown in figure 2
\cite{bonato1,bonato2,simon1}. Each branch contains a 4f imaging
system with lenses of focal length $f$ and a thin object that
provides spatial modulation $G_j({\mathbf y})$ of the beam, where
$j=1,2$ labels the branch, and ${\mathbf y}$ is the position in
the plane transverse to the axis. The goal is to cancel
object-induced optical aberrations (position-dependent phase
shifts produced by the $G_j$). The case of a single object in one
branch only may be included by simply setting $G=1$ in the other
branch. The plane of the samples (labelled $\Pi$ in fig. 2) is the
Fourier plane of the 4f system. Time delay $\tau$ is inserted in
one branch. In the detection stage, two large bucket detectors
$D_1$ and $D_2$, connected in coincidence, record the arrival of
photons, but not their positions. Apertures described by pupil
functions $p_1({\mathbf x}_1)$ and $p_2({\mathbf x}_2)$ are
followed by crossed polarizers at $45^\circ$ to each beam's
polarization, before arriving at the detectors.

\begin{figure}
\centering
\includegraphics[totalheight=2.0in]{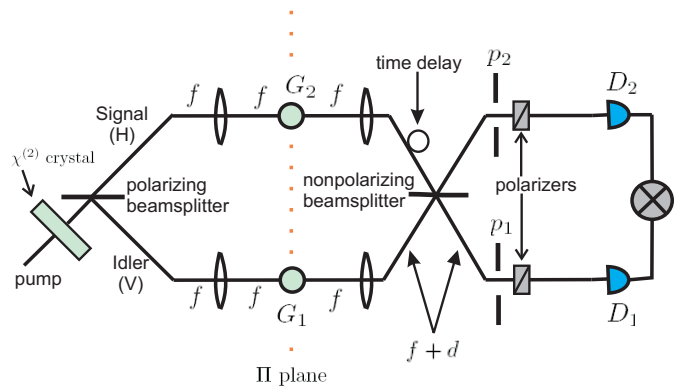}
\caption{\textit{(Color online) Schematic of interferometer with even-order
aberration-cancellation. Large bucket detectors $D_1$ and $D_2$
are integrated over and connected by a coincidence circuit.}}
\end{figure}

A continuous wave laser pumps a $\chi^{(2)}$ nonlinear crystal,
leading to collinear type II parametric downconversion. The
frequencies of the two photons are $\Omega_0\pm \nu$, with
transverse momenta $\pm {\mathbf q}$. For simplicity, assume the
frequency bandwidth is narrow compared to $\Omega_0$. The two
photons have total wavenumbers ${{\Omega_0\pm \nu}\over c}\approx
{{\Omega_0}\over c}$. The downconversion spectrum is
\begin{equation}\Phi({\mathbf q},\nu )= sinc\left[ {{L\Delta ({\mathbf q},\nu )}\over 2}\right] e^{i{{L\Delta ({\mathbf q},\nu )}\over 2}}
\label{spectrum},\end{equation} where $L$ is the thickness of the
crystal, and
\begin{equation} \Delta ({\mathbf q},\nu ) =-\nu D +M\hat {\mathbf e_2}\cdot
{\mathbf q}+{{2|{\mathbf q}|^2}\over {k_{pump}}} .\end{equation}
$D$ is the difference between the inverse group velocities of the ordinary
and extraordinary waves in the crystal, and $M$ is the spatial
walk-off in the direction $ \hat{{\mathbf e_2}}$ perpendicular to
the interferometer plane. The last term in $\Delta$ is due to
diffraction in the crystal. Ignoring the vacuum term and terms of
higher photon number, the wavefunction entering the apparatus is
approximately given by
\begin{eqnarray}|\Psi\rangle &=&\int d^2q\; d\nu\; \Phi({\mathbf q},\nu )\nonumber \\
& & \times \; \hat
a_s^\dagger ({\mathbf q},\Omega_0+\nu )\hat a_i^\dagger (-{\mathbf q},\Omega_0-\nu )
|0\rangle ,\end{eqnarray} where $\hat a_s$ and $\hat a_i$ are
annihilation operators for the signal and idler photons. For
collinear pairs, horizontally polarized photons are directed into
the upper branch and vertically polarized photons into the lower
branch by means of a polarizing beamsplitter. Alternatively,
noncollinear pairs could be used with polarizers selecting
horizontal (H) polarization in the upper branch and vertical (V)
in the lower one. In either case, we will refer to the H photon in
the upper branch (branch 2) as the signal and the V photon in
branch 1 as the idler.

The coincidence rate is of the generic form \cite{rubin}
\begin{equation} R(\tau )=R_0\left[
1-\Lambda \left( 1-{{2\tau}\over {DL}}\right) W(\tau )\right] ,
\end{equation} where $\Lambda (x)$ is the triangular function:
\begin {equation}
\Lambda (x) = \bigg \{
\begin{array}{l}
1 - |x|, \qquad |x| \leq 1 \\
0, \qquad \qquad |x| > 1
\end{array} \end{equation}
For large apertures, $p_1({\mathbf x}_1)=p_2({\mathbf x}_2)\approx
1$; so, as shown in \cite{bonato2}, the background and
$\tau$-modulation terms are
\begin{eqnarray} R_0&=& \int d^2q
\left| G_1 \left( {{f{\mathbf q} }\over k}\right) G_2 \left( -{{f{\mathbf q} }\over
k}\right) \right|^2 \label{tau}\label{background2}\\
W(\tau )&=& {1\over {R_0}}\int d^2q e^{-{{2iM\tau}\over D}{\mathbf
e_2}\cdot {\mathbf q}}\label{mod2}\\ & &\; \times \; G_1^\ast \left(
{{f{\mathbf q} }\over k}\right)G_1 \left( -{{f{\mathbf q} }\over
k}\right)\nonumber \\
& & \times\; G_2^\ast \left( -{{f{\mathbf q} }\over k}\right)G_2
\left( {{f{\mathbf q} }\over k}\right)\nonumber ,\end{eqnarray}
where $k$ is the longitudinal wavenumber.

We may write $G_j({\mathbf x})=t_j({\mathbf x})e^{i\phi_j({\mathbf
x})}$, with $t_j$ real. Aberration effects arise from
spatially-dependent phase factors $\phi_j({\mathbf x})$, which
lead to distortion of the outgoing wavefronts. The phase functions
may be decomposed into a sum of pieces that are either even under
reflection, $\phi_j^{(even)}(-{\mathbf
x})=\phi_j^{(even)}({\mathbf x})$ or odd,
$\phi_j^{(odd)}(-{\mathbf x})=-\phi_j^{(odd)}({\mathbf x})$.
Astigmatism and spherical aberration, for example, are included in the
even-order part, whereas coma is odd.

In equation (\ref{mod2}), the factors $G_1^\ast \left( {{f{\mathbf
q} }\over k}\right)G_1 \left( -{{f{\mathbf q} }\over k}\right)$
become
\begin{equation} t_1^\ast \left( {{f{\mathbf q}
}\over k}\right)t_1 \left( -{{f{\mathbf q} }\over
k}\right)e^{-i\left[\phi_1\left( {{f{\mathbf q} }\over
k}\right)-\phi_1\left(-{{f{\mathbf q} }\over k}\right)\right]
} .\label{phases}\end{equation} the form of the difference in the
exponent shows that even-order aberrations arising from object 1
cancel from the modulation term. The even-order aberrations from
object 2 cancel in a similar manner. This is the even-order
cancellation effect demonstrated in \cite{bonato1} and
\cite{bonato2}. As pointed out in \cite{simon1}, even- and
odd-orders cancel simultaneously only in the special case
$G_1=G_2$. These cancellations are exact only for aberrations
induced by thin objects in the particular plane $\Pi$.

In the term $R_0$, both even-order and odd-order aberrations
cancel even for $G_1\ne G_2$. For time correlation experiments,
this is an unimportant background term; however this term is the
foundation of the imaging apparatus described below, since the
beamsplitter will be absent, meaning that there will be no
modulation term $W(\tau )$. The physical mechanism of the various
possible cancellations are discussed in more detail in
\cite{simon1}.

\section{ABERRATION-CANCELLED GHOST IMAGING WITH ENTANGLED SOURCE} Now
we wish to look at the ghost imaging analog of the aberration-cancelling
interferometer of the previous section. This leads us to
the hybrid device of fig. 3.  This
new apparatus differs from that of fig. 2 in several respects.
First, we have removed the time delay, polarization filters, and
beam splitter; these were needed to produce the interference
effects desired in \cite{bonato1,bonato2,simon1}, but are not
necessary for imaging purposes. Also, in order to obtain spatial
resolution, one bucket detector ($D_1$) is replaced by a moveable
pointlike detector or a CCD camera. The removal of the beam
splitter and the introduction of spatial resolution are the key
changes. After allowing for an arbitrary source of correlated
(quantum or classical) light, we arrive at an apparatus in fig. 3
that looks very much like the ghost imaging setup of fig. 1, but
with a 4f imaging system in each branch. In this section, we
assume that the light source is parametric downconversion.

\begin{figure}
\centering
\includegraphics[totalheight=2.0in]{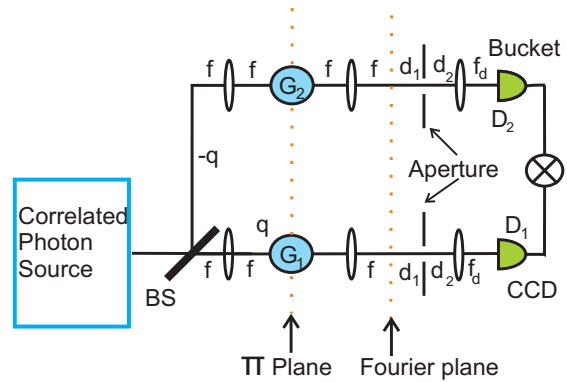}
\caption{\textit{(Color online) Schematic of correlated-photon imaging setup with
aberration-cancellation. All orders of aberration cancel. }}
\end{figure}

The coincidence rate at location ${\mathbf x_1}$ of $D_1$ is
\begin{equation} R({\mathbf x_1}) = \int d^2x_2dt_1dt_2 |A({\mathbf x_1},{\mathbf x_2},t_1,t_2)|^2
,\end{equation} where the transition amplitude is \begin{equation}
A({\mathbf x}_1,{\mathbf x}_2,t_1,t_2) = \langle
0|E_1^{(+)}({\mathbf x}_1,t_1) E_2^{(+)}({\mathbf x}_2,t_2) |\Psi
\rangle \label{amplitude}.\end{equation}

Taking the two detection apertures described by $p_1$ and $p_2$ to
be large, we compute the coincidence rate to be
\begin{eqnarray} R({\mathbf x_1})&=& \left\{ \left[ B({\mathbf x_1})+B(-{\mathbf x_1})\right]
+\left[ C({\mathbf x_1})+C^\ast ({\mathbf x_1})\right]\right\} \nonumber \\
& & \quad \times \; \left| G_1\left( {f\over {f_D}}{\mathbf x_1}\right)
G_2\left( -{f\over {f_D}}{\mathbf x_1}\right) \right|^2 ,\end{eqnarray}
where
\begin{eqnarray} B({\mathbf x_1})&=&\int d\nu \left| \Phi
\left( {{k{\mathbf x_1}}\over {f_D}} ,\nu \right) \right|^2 ,\\
C({\mathbf x_1})&=&\int d\nu \; \Phi \left( {{k{\mathbf x_1}}\over {f_D}} ,\nu \right)
\Phi^\ast \left( -{{k{\mathbf x_1}}\over {f_D}} ,-\nu \right) .
\end{eqnarray}
Using eq. (\ref{spectrum}), these integrals may be evaluated; they
turn out to be $x_1$-independent constants. Sweeping all overall
constants into a single constant ${\cal R}_0$, we find:
\begin{equation} R({\mathbf x_1}) = {\cal R}_0  \left| G_1\left(
{f\over {f_D}}{\mathbf x_1}\right) G_2\left( -{f\over
{f_D}}{\mathbf x_1}\right) \right|^2 . \label{Rx}\end{equation}
Only the modulus of each $G_j$ enters into $R({\mathbf x})$, so we
see that the aberrations introduced by the object phases cancel to
all orders. This will be exact only in the Fourier plane, but as
was true for the interferometer case, we would expect it to continue to remain approximately true
as we move out of the plane up to a maximum distance on the order of ${{fr_s}\over a}$, where
$f$ and $a$ are the focal length and radius of the lens and $r_s$ is the maximum radius of
the object being viewed. (See ref.(\cite{simon1}) for a derivation of this estimate.)

If $G_2=1$, then we have an ordinary (non-ghost) image
of $\left| G_1\right| $. On the other hand, if $G_1=1$ then we
have an inverted ghost image of $\left| G_2\right|$. In either
case, the image is magnified by a factor of $m={{f_D}\over f}$. Note
that, in contrast to the interferometry case, even and odd order phases
both cancel, even in the general case $G_1\ne G_2$. The cancellation of the phases only occurs in the Fourier plane; aberrative distortions begin
growing when the objects are moved out of this plane.

\section{APPLICATIONS TO IMAGE ANALYSIS}

Note that if both $G_1$ and $G_2$ are nontrivial objects, what we actually see is their pointwise product. This is a distinctive feature of this apparatus. It can be verified by straightforward calculation that the simple product structure of eq. (\ref{Rx}) does not occur in other obvious variations of two-object ghost imaging systems; for example, it does not occur if the 4f imaging system in either branch (or both) is replaced by a single lens imaging system, or in a system
without lenses.
We may make use of this product structure in a number of ways. For example, if the object of interest is $G_2$ but the second branch
introduces some {\it known} distortion to its image (e.g. there may be aberration in the optical system or
 variations in the refractive index of the propagation medium), then this can be cancelled by using an object $G_1$ that introduces an opposite distortion (via a deformable mirror, for example). The two distortions then cancel, leaving no net effect on the image. Alternatively, if $G_2$ has a dim, low-transmissivity area that we wish to view, but it is being obscured by a bright, high-transmissivity area nearby, we may use a mask
for $G_1$ which allows a view only of the twins of photons coming from the dim region of interest, blocking
photons that are partnered with light from other areas.

At first glance, the product structure might seem to open up a further interesting possibility. Suppose $G_1$ is the object we wish to view. If there is no object in branch 2 ($G_2$ is simply a constant), then the resolution with which we may view $G_1$ is the same as if the second branch was not there. It would be limited by the
sizes of the Airy disks produced by the lenses. However, if $G_2$ is taken to be a small pinhole, the area we would be able to see of $G_1$ at any given time would be limited by the size of the pinhole. Thus, it would seem that by taking the pinhole small
enough, we would be able to limit our view of $G_1$ to an area smaller than the standard Abb\'e limit, thus achieving subresolution imaging. Unfortunately, when the finite sizes of the lenses are properly taken into account (eq. (\ref{Rx}) was derived in the limit of large lenses), the combined action of diffraction in the two branches conspires to give the single-branch resolution as its best-case limit, occurring when the pinhole radius is negligible. As the pinhole radius at $G_2$ grows to finite size, the resolution becomes worse than in the single-branch case.

One additional observation on applications of the product structure arises if we replace the position-resolving detector in branch 1 by a bucket detector, thus introducing an integration over ${\mathbf x_1}$. We have now lost all imaging ability, but note what happens if we displace one of the objects (object 1, say) by some distance in the transverse plane. If the 2-dimensional displacement vector is ${\mathbf r}$, then equation (\ref{Rx}) is replaced by \begin{equation} R({\mathbf r}) = {\cal R}_0 \int \left| G_1\left(
{f\over {f_D}}({\mathbf x_1}+{\mathbf r})\right) G_2\left( -{f\over
{f_D}}{\mathbf x_1}\right) \right|^2 dx_1 .\label{cor} \end{equation}
Thus, despite the fact that neither detector has spatial resolution, the system optically computes the spatial intensity correlator $g({\mathbf r})=\langle I_1 \left( m^{-1}({\mathbf x}+{\mathbf r})\right) I_2\left( -m^{-1}{\mathbf x})\right)\rangle $, where $m$ is the magnification. (The correlation here is actually between the object $G_1({\mathbf x})$ and the {\it inverted} object $G_2(-{\mathbf x})$, but an additional lens can be added
to remove remove the inversion and cancel the minus sign in $G_2$.)
The full correlation function can be found by moving one object repeatedly to scan over the full range of relevant ${\mathbf r}$ vectors. Taking one of the two objects to be unknown and the other to be some known template, this could provide a means of identifying the unknown object by quantifying its degree of similarity to the template. This could be useful, for example, in comparing silicon chips on an assembly line to a standard chip, and identifying those chips with flaws. Note in particular, that the unknown object may be in a remote, inaccessible location; for example, the object might be a cell inside the body being viewed through an endoscope and compared to a cell in the lab. As in the case of the temporal correlator studied with the interferometer of refs. \cite{bonato1,bonato2,simon1}, the effect of object-induced aberrations (differences between phase shifts induced by the two samples) cancels out of the spatial correlator.


\section{IMAGING WITH A CLASSICAL SOURCE} We
now replace the downconversion source of the previous section
by a classical source of anticorrelated photons, as in
\cite{bennink1}. Light entering a beam splitter with transverse
momentum ${\mathbf q}$ leads to outgoing beams with
anticorrelated momenta ${\mathbf q}$ and $-{\mathbf q}$. If the
beam steering modulator produces momentum spectrum $f({\mathbf
q})$, the input state for pairs of photons having the same
${\mathbf q}$ before the beamsplitter will be $\sim \int d^2q
F({\mathbf q}) \hat a_p^\dagger ({\mathbf q})\hat a_p^\dagger
({\mathbf q})|0\rangle $, where $\hat a_p^\dagger$ is the creation
operator for pump photons and $F({\mathbf q})\equiv f^2({\mathbf
q})$. We assume for simplicity that $F({\mathbf q})$ is an even
function, $F({\mathbf q})=F(-{\mathbf q})$. Denoting creation
operators in the two outgoing branches by $\hat a_1^\dagger$ and
$\hat a_2^\dagger$, the incoming photon pair will produce a state
after the beamsplitter given by: \small{\begin{eqnarray} |\Psi
\rangle &=& {1\over 2} \int d^2q F({\mathbf q}) \left[ \hat
a_1^\dagger ({\mathbf q})+\hat a_2^\dagger (-{\mathbf q})\right] \\
& & \qquad \qquad \qquad\qquad
\times \left[ \hat a_1^\dagger ({\mathbf q})+\hat a_2^\dagger
(-{\mathbf q})\right] |0\rangle \nonumber \\
&=&\int d^2q F({\mathbf q}) \left[ \hat a_1^\dagger ({\mathbf
q})\hat a_2^\dagger (-{\mathbf q})+\; \dots\; \right]
|0\rangle ,\label{classstate}\end{eqnarray}} where the terms
dropped in the last line are those which do
not contribute to coincidence detection. The detection
amplitude of eq. (\ref{amplitude}) is then proportional to
\begin{equation} \int d^2q
F({\mathbf q})e^{i{\mathbf q}\cdot ({\mathbf x_1}-{\mathbf x_2})}
H_1({\mathbf q},{\mathbf x_1})H_2(-{\mathbf q},{\mathbf
x_2})\label{classamp},\end{equation} where $H_j$ is the
transfer function for branch $j$. Integrating over $D_2$, we then
have the coincidence rate: \begin{equation} R({\mathbf x_1}) =
\left| F\left( {k\over {f_D}}{\mathbf x_1}\right) G_1\left(
{f\over {f_D}}{\mathbf x_1}\right)G_2\left( -{f\over
{f_D}}{\mathbf x_1}\right)\right|^2
.\label{classrate}\end{equation} This is similar to the result for
the entangled-source apparatus, except modulated by the factor
$F\left( {k\over {f_D}}{\mathbf x_1}\right)$ which is determined
by the details of the beam steering modulator. Similarly, for
thermal or speckle sources, this factor will arise from the
transverse momentum spectrum of the source.

\section{ROLE OF THE DETECTION LENS} Consider
now the lenses immediately before the detectors in fig.
3. With no such detection lens present, the transfer function for
branch $j$ would be
\begin{equation} H_j({\mathbf q_j},{\mathbf x_j} )=
G_j\left({{f{\mathbf q_j} }\over k}\right) e^{i{\mathbf q}\cdot {\mathbf
{\mathbf x_j}}} ,\label{nolens} \end{equation} from which we see that the
information from each ${\mathbf q}$ value is spread over all ${\mathbf x}$
values. But {\it with} the lens, eq. (\ref{nolens})
becomes
\begin{eqnarray}  H_j({\mathbf q_j},{\mathbf x_j} )&=&
 e^{-{{ik{\mathbf
x_j}^2}\over {2f_D}}\left( {{d_2 }\over {f_D}}-1\right)}e^{-{{id_1
{\mathbf q_j}^2}\over {2k}}} \nonumber \\
& & \times\; G_j\left( {{f{\mathbf q_j} }\over k}\right)\delta \left({{k{\mathbf x_j}}\over
{f_D}}-{\mathbf {\mathbf q_j}}\right)\label{lens},\end{eqnarray}
so that each ${\mathbf q}$ value is localized at a single point in
the detector plane via the delta function. Since each ${\mathbf
q}$ value is also matched to an object point, the localization in
the second case defines a mapping between points in the object
plane and points in the detection plane, allowing reconstruction
of an image by the pointlike detector $D_1$. This can be verified
by computing the coincidence rate with or without the final
lenses, i.e. using either eq. (\ref{nolens}) or eq. (\ref{lens}).
Doing so, we find that without the branch 1 lens the coincidence
rate becomes independent of ${\mathbf x_1}$, making imaging
impossible. In contrast, removing the branch 2 lens has no effect.
This makes intuitive sense: we integrate over ${\mathbf x_2}$, so
it does not matter if the momentum information in this branch is
localized or spread over the entire detector. Thus we arrive at an
important technical point: the lens before the bucket detector may
be removed without harm, but the branch 1 lens is essential for
imaging.

The need for a lens before $D_1$ may be viewed as follows. The 4f
system in either branch transfers modulation $G_j$ from the
transverse coordinate space (${\mathbf x}$) to the Fourier space
(${\mathbf q}$), which is where the aberration cancellation
actually takes place (see \cite{simon1}). The lens in front of
$D_1$ is then needed to transfer the modulation back to coordinate
space for imaging.

\section{CONCLUSIONS} In conclusion, we have proposed a new type
of two-object ghost imaging apparatus that cancels phase
effects from thin objects in the vicinity of
a particular plane and which allows comparisons between pairs of objects.
The method involves a relatively simple apparatus and
can be done with either entangled photon pairs or with a
classically-correlated light source. This apparatus
may have potential for new applications in biomedical research, industry,
and other fields.

\section*{ACKNOWLEDGEMENTS}
This work was supported by a U. S. Army Research Office (ARO)
Multidisciplinary University Research Initiative (MURI) Grant; by
the Bernard M. Gordon Center for Subsurface Sensing and Imaging
Systems (CenSSIS), an NSF Engineering Research Center; by the
Intelligence Advanced Research Projects Activity (IARPA) and ARO
through Grant No. W911NF-07-1-0629.

The authors would like to thank Dr. Lee Goldstein and Dr. Robert Webb for some very useful discussions
and advice.

\end{document}